\documentclass[11pt]{article}
\usepackage{cospar}
\usepackage{url}
\usepackage{natbib}
\usepackage{times}

\usepackage{graphicx}
\usepackage{floatflt}
\usepackage[figuresright]{rotating}

\newcommand{\pub}[5] {#5, {\it #1}, {\bf #2}, #3, #4.}
\newcommand{\puba}[5]{#5, {\it #1}, #3, #4.}

\newcommand{\placefigure}[1]{}
\newcommand{\placetable}[1]{}

\newcommand{\antiproton}{$\bar p$}
\newcommand{\proton}{$p$}
\newcommand{\Dxx}{D_{xx}}
\newcommand{\e}[2]{$^{#1}_{#2}$}
\newcommand{\twofigs}{0.48\textwidth}
\newcommand{\two}{0.5\textwidth}

\newcommand\paperno{
   \vspace{-16\baselineskip} \hspace{-0.05\textwidth}
   \begin{minipage}[t]{100mm}
   \noindent \bf H0.2-E0.2-0035-02 \rm(Proc.~COSPAR-2002, Houston, TX)\\
\bf LANL Report \# LA-UR-02-7813\\
\small \rm Submitted to \it Advances in Space Research
   \end{minipage}\vspace{14\baselineskip}}
\newcommand{\aap}{Astron.\ Astrophys.}
\newcommand{\adv}{Adv.\ Space Res.}
\newcommand{\apj}{Astrophys.\ J.}
\newcommand{\apjl}{Astrophys.\ J.}

\newcommand{\mnras}{Mon.\ Not.\ Royal Astron.\ Soc.}
\newcommand{\plb}{Phys.\ Lett.\ B}
\newcommand{\prc}{Phys.\ Rev.\ C}
\newcommand{\prl}{Phys.\ Rev.\ Lett.}
\newcommand{\ssr}{Spa.\ Sci.\ Rev.}

\title{PROPAGATION OF SECONDARY ANTIPROTONS AND COSMIC RAYS IN THE GALAXY}

\author{I. V. Moskalenko\address{NASA/Goddard Space Flight Center,
        Code 661, Greenbelt, MD 20771, U.S.A.}\address{Joint Center for 
        Astrophysics, University 
        of Maryland, Baltimore County, Baltimore, MD 21250, U.S.A.},
        A. W. Strong\address{MPI f\"ur extraterrestrische Physik, 
        Postfach 1603, D-85740 Garching, Germany},
        J. F. Ormes$^1$,
        and
        S. G. Mashnik\address{Los Alamos National Laboratory, Los Alamos,
        NM 97545, U.S.A.}}

\begin{document}

\maketitle
\paperno

\begin{abstract}
Recent measurements of the cosmic ray (CR) antiproton flux
have been shown to challenge existing CR propagation
models. 
It was shown that the reacceleration models designed to match
secondary to primary nuclei ratios (e.g., B/C) produce too few
antiprotons. 
In the present paper we discuss one possibility to overcome these
difficulties. Using the
measured antiproton flux \emph{and}
B/C ratio to fix the diffusion coefficient, we show that 
the spectra of primary nuclei as measured in the heliosphere may contain a
fresh local ``unprocessed'' component at low energies
perhaps associated with the Local Bubble, thus decreasing
the measured secondary to primary nuclei ratio.
The independent evidence for SN activity in the solar vicinity
in the last few Myr supports this idea.
The model reproduces
antiprotons, B/C ratio, and elemental abundances up to Ni ($Z\leq28$).
Calculated isotopic distributions of Be and B 
are in perfect agreement with CR data.
The abundances of three ``radioactive clock'' isotopes in CR, \e{10}{}Be,
\e{26}{}Al, \e{36}{}Cl,
are all consistent and indicate a halo size $z\,_h\sim4$ kpc based on 
the most accurate data taken by the ACE spacecraft.

\end{abstract}

\section*{INTRODUCTION} \label{sec:intro}

The result of our recent analysis \citep{M01,M02}, in agreement with calculations of
other authors \citep{simon}, was that matching the
secondary/primary nuclei ratio B/C using reacceleration models leads
to values of the spatial diffusion coefficient apparently too large to produce
the required \antiproton\ flux, when the propagated nucleon spectra are
tuned to match the local \proton\ and He flux measurements.
Assuming the current heliospheric modulation models are approximately
right, we have the following possibility \citep{M02b} 
to reconcile the B/C ratio with the
measured flux of secondary \antiproton's.  The spectra of
primary nuclei as measured in the heliosphere may contain a fresh
local ``unprocessed'' component at low energies thus decreasing the
measured secondary to primary nuclei ratio.  
This component would have to be local in the sense of being 
specific to the solar neighbourhood, so that the well-known 
``Local Bubble'' phenomenon
is a natural candidate.

The low-density region around the Sun, filled with hot H~{\sc i} gas,
is called the Local Bubble (hereafter LB) \citep[e.g.,][]{sfeir}. 
The size of the region is about 200 pc, 
and it is likely that it was produced in a series of supernova
(SN) explosions, with the last SN explosion occuring approximately 
1--2 Myr ago, or 3 SN occuring  within the last 5 Myr. 
Most probably its progenitor was an OB 
star association \citep[e.g.,][]{maiz,berghofer}.
\citet{fe60} report about an excess of 
$^{60}$Fe measured in a deep ocean core sample of
ferromanganese crust suggesting 
a SN explosion about 5 Myr ago at 30 pc distance
and the deposition of SN-produced iron on earth.
\citet{sonett} report an enhancement in the CR intensity dated about 40 kyr
ago, which is interpreted as the passage 
across the solar system of the shock
wave from a SN exploding about 0.1 Myr ago.
It could also be that ``fresh'' LB contributions from
continuous acceleration in the form of shock waves \citep{bykov},
and/or
energetic particles coming directly from SN remnants still influence the spectra and
abundances of local CR.

\section*{THE CALCULATION PROCEDURE}

We use the propagation model 
GALPROP\footnote{GALPROP model including software and data sets is 
available at {http://www.gamma.mpe-garching.mpg.de/$\sim$aws/aws.html}}
as described elsewhere \citep{SM98,M02}; for the present purpose 
the 2D cylindrically symmetrical option is sufficient. 
The spatial diffusion coefficient is
taken as $\Dxx = \beta D_0(\rho/\rho_0)^{\delta}$;  
the diffusion in momentum space and other details 
can be found in our earlier papers. The nucleon injection spectrum of the
Galactic component was taken as a modified power-law in rigidity
\citep{jones}, $dq(p)/d\rho \propto
\rho^{-\gamma}/\sqrt{1+(\rho/2)^{-2}}$, for the injected particle density.
The heliospheric modulation is
treated using the force-field approximation.
The LB spectrum is taken in the form 
\citep[][for continous acceleration by interstellar shocks]{bykov}:
$df/d\rho \propto \rho^{-\eta}\exp(-\rho/\rho_b)$, where $\rho$ is the
rigidity, and $\rho_b$ is the cut off rigidity parameter.
In terms of kinetic energy per nucleon $E$ this can be re-written as
$df/dE= a(Z,A)\frac{A(E+m)}{Zp}\rho^{-\eta}e^{-\rho/\rho_b},$
where $a(Z,A)$ is the abundance of a nucleus $(Z,A)$, $Z,A$ are the nucleus
charge and atomic number, correspondingly, $m$ is the atomic mass unit,
$p$ is the momentum per nucleon, $\rho\,_b=\frac{A}{Z}\sqrt{(E_b +m)^2 -m^2}$.
We show the results obtained with $\eta=1$, but the particular spectral shape of the LB
component is not important as long as it decreases sharply towards high
energies and is much softer than the Galactic CR spectrum.  

\begin{figure*}[tb] 
\includegraphics[width=\two]{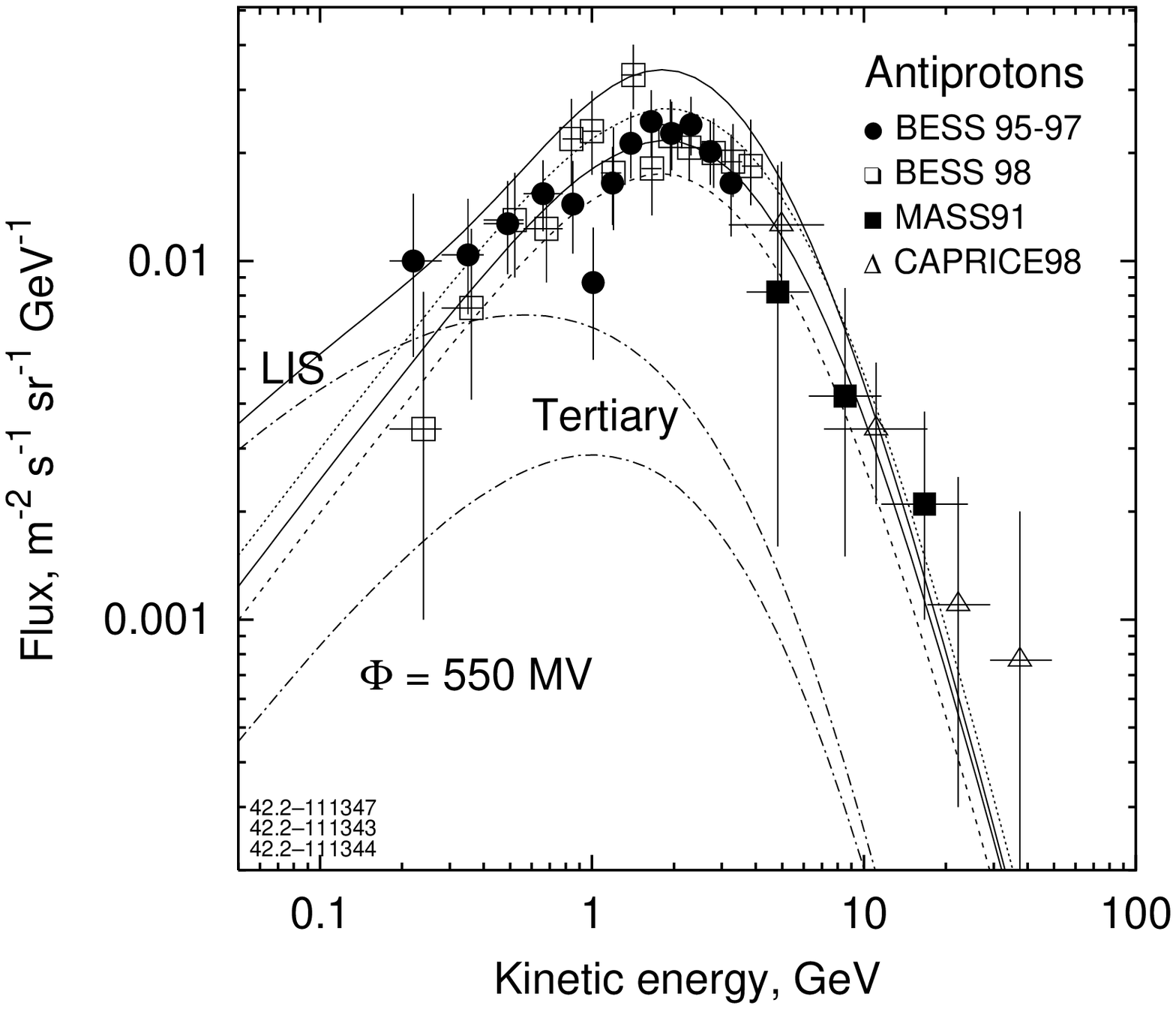}\hfill
\includegraphics[width=\two]{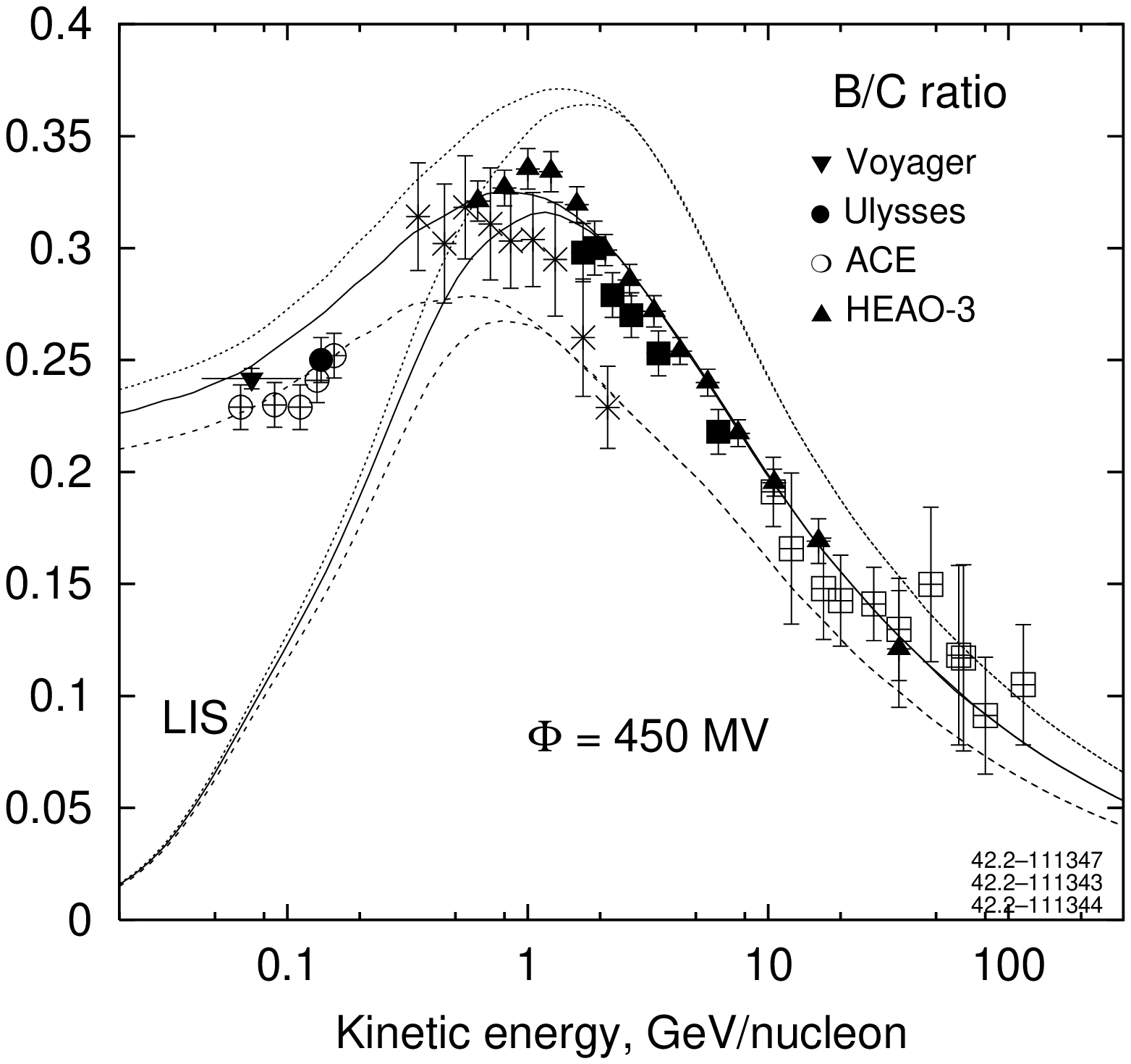}
\vspace{-3\baselineskip}
\caption{\small
\emph{Left:} Calculated \antiproton\ flux in a model 
with $\delta=0.47$ and different normalization
values $D_0$ ($\times10^{28}$ cm s$^{-2}$).
Solid curves -- $D_0=3.3$ at $\rho_0=3$ GV, 
upper curve -- local interstellar (LIS), 
lower curve -- modulated ($\Phi=550$ MV). 
Dots --  $D_0=2.6$, 
dashes -- $D_0=4.3$. 
Data: BESS 95-97 \citep{Orito00}, BESS 98 \citep{Asaoka02}
MASS91 \citep{Stoc01}, CAPRICE98 \citep{boezio01}.
\emph{Right:} B/C ratio calculated with LB contribution,
$E_b=500$ MeV. The lines are coded as on the left panel.
Lower curves -- LIS, upper -- modulated ($\Phi=450$ MV).
Data below 200 MeV/n: ACE \citep{davis}, Ulysses
\citep{ulysses_bc}, Voyager \citep{voyager}; high energy data:
HEAO-3 \citep{Engelmann90}, for other references see
\citet{StephensStreitmatter98}.\vspace{-0.5\baselineskip}
\label{fig:pbars}}
\end{figure*}

The procedure to tune the CR elemental abundances, secondary/primary
nuclei ratios, and \antiproton\ flux we adopted was as follows.
The high energy part of B/C ratio \emph{and} \antiproton\ flux measurements
are used to restrict the parameters of the diffusion coefficient $D_0$ and 
$\delta$, while the low energy part of B/C ratio is used
to fix a value for the reacceleration level and define 
the parameters of the LB component.
The Galactic CR elemental source abundances are tuned
using the CR abundances at high
energies where the heliospheric modulation is weak.
The LB isotopic abundances are tuned to
match the low-energy data by ACE and Ulysses.

Figure~\ref{fig:pbars} (left) illustrates the process of fixing the
normalization of the diffusion coefficient using \antiproton's. 
The \antiproton\ flux shown is calculated with
$\delta=0.47$ and different normalizations in the diffusion coefficient,
$D_{\,0}=2.6, 3.3, 4.3\times10^{28}$ cm s$^{-2}$ at $\rho=3$ GV 
(for \antiproton's this corresponds to 
kinetic energy $\sim2$ GeV). The injection index $\gamma$ is taken
equal to 2.28, and the Alfv\'en speed $v_A=23$ km s$^{-1}$.
The \antiproton\ flux at maximum, $\sim2$~GeV, appears
to be quite sensitive to the value of the diffusion coefficient and
allows us to fix it at $D_{\,0}=(3.3\pm0.8)\times10^{28}$ cm s$^{-2}$.
The exact value of $\delta$ is not critical since we compare with
the \antiproton\ measurements at maximum.
Inelastically scattered \antiproton's, the ``tertiary'' component,
appears to be important at low energies only in the ISM.
Figure\ \ref{fig:pbars} (right) shows corresponding calculations of the B/C ratio.
A halo height of $z_h=4$ kpc is used \citep{SM01,MMS}. 

The local interstellar (LIS) \proton\ and He spectra used in our 
\antiproton\ calculations are the best fits
to the data as described in \citet{M02}. 
Because of the measurements with large statistics (BESS --
\citealt{Sanu00}, AMS -- \citealt{p_ams}), and weak 
heliospheric modulation above 10 GeV, 
the error arising from 
uncertainties in the primary spectra 
is only $\sim5$\%. 
We note that as in the case of other nuclei, there
should be an LB contribution to \proton\ and He spectra.
The Galactic injection
spectra of \proton's and He should thus be significantly flatter below several GeV
to match the data points at low energies.
This does not influence the \antiproton\ production because
(i) the LB does not produce a significant amount of secondaries, and (ii)
the \antiproton\ threshold production energy is high, $\sim10$ GeV.

Further tuning can be done
using the high energy part of the B/C ratio, which is not influenced
by heliospheric modulation and supposedly contains only a Galactic
component of CR. Index $\delta\sim0.47$ 
is chosen as giving the best match. A pure Kolmogorov spectrum,
$\delta=1/3$, thus seems to be excluded by \antiproton\ spectrum data 
taken in combination with B/C ratio data.
The low energy B/C data were used to fix the LB spectral parameter at
$E\,_b=500$ MeV, but $E_b=400-600$ MeV also provides good agreement with B/C data.
In this way, we have been able to obtain a model
simultaneously fitting \proton, He, \antiproton, and B/C data.

\section*{ABUNDANCES IN COSMIC RAYS AND COSMIC RAY SOURCES} \label{sec:results}

The Galactic CR source elemental abundances are tuned (at a nominal 
reference energy of 100 GeV), by a least squares procedure, to
the abundances measured by HEAO-3 \citep{Engelmann90} at 5.6, 7.5, 10.6,
16.2 GeV/n.
Relative isotopic abundances at the source are taken
equal to solar system abundances \citep{solar_isot_ab}.
The key point in the fitting procedure is to obtain the correct 
abundance of B.

\begin{floatingfigure}[r]{\twofigs}
\sf
\begin{center}
\includegraphics[width=\twofigs]{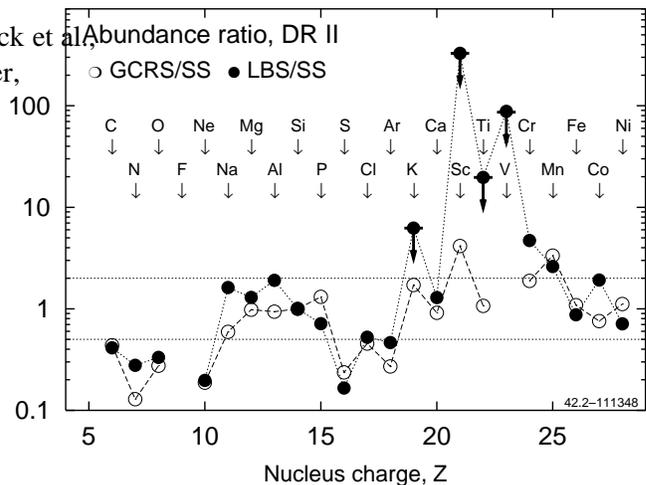}
\end{center}
\caption{\small 
Derived abundance ratios
Galactic-CR-source to solar-system (GCRS/SS) 
and LB-source to solar-system (LBS/SS), normalized to Si. Relative
abundances for K, Sc, Ti, V are shown as upper limits.
The dotted lines plotted at 1/2 and 2.\vspace{0.3\baselineskip}
\label{fig:Ab-ratio}}
\end{floatingfigure}

The LB elemental
abundances are tuned simultaneously with spectra using the
low energy part of the B/C ratio and isotopic abundances at 200 MeV/n 
from ACE \citep{ace_data} and at 185 MeV/n from Ulysses \citep{ulysses}.
For many elements ACE and Ulysses abundances differ by 10\%.
For this reason, to the statistical
errors we added 5\% systematic error. 
The fitting procedure is influenced by the adopted value 
of the modulation potential.
The following pairs of modulation potentials give almost the same
final result: (ACE, Ulysses) = (400 MV, 700 MV), (450 MV, 600 MV),
(500 MV, 500 MV).
The data deviate from calculations in
both directions, which mean that we are unlikely to
introduce essential systematic error by assuming a wrong value of the
modulation potential.

The diffusive reaccleration model with an LB component shows good overall 
agreement with data including secondary to primary ratios, spectra, 
and abundances. 
The derived Galactic CR source abundances and LB source abundances are
plotted in Figure~\ref{fig:Ab-ratio} relative to the solar system abundances
\citep{solar_elem_ab}.  
The important result is that CR source
and LB source abundances of all major elements 
(C, O, Ne, Mg, Si, S, Ar, 
Ca, Fe, Ni) are in good agreement with each other.
Abundances of the Si group, Ca, Fe, Ni are near the 
solar system abundances.
Abundances of other elements in Galactic CR and LB sources 
are mostly consistent with each other, and with
solar system abundances, within a factor of 2.
Relative to silicon, C, N, O, Ne, S 
are underabundant both in CR source and LB. 
This corresponds to the well-known 
first-ionization-potential or
volatily correlation \citep[see, e.g.,][]{meyer}. 

\begin{figure}[tb]
\footnotesize
\begin{center}
\includegraphics[width=\two]{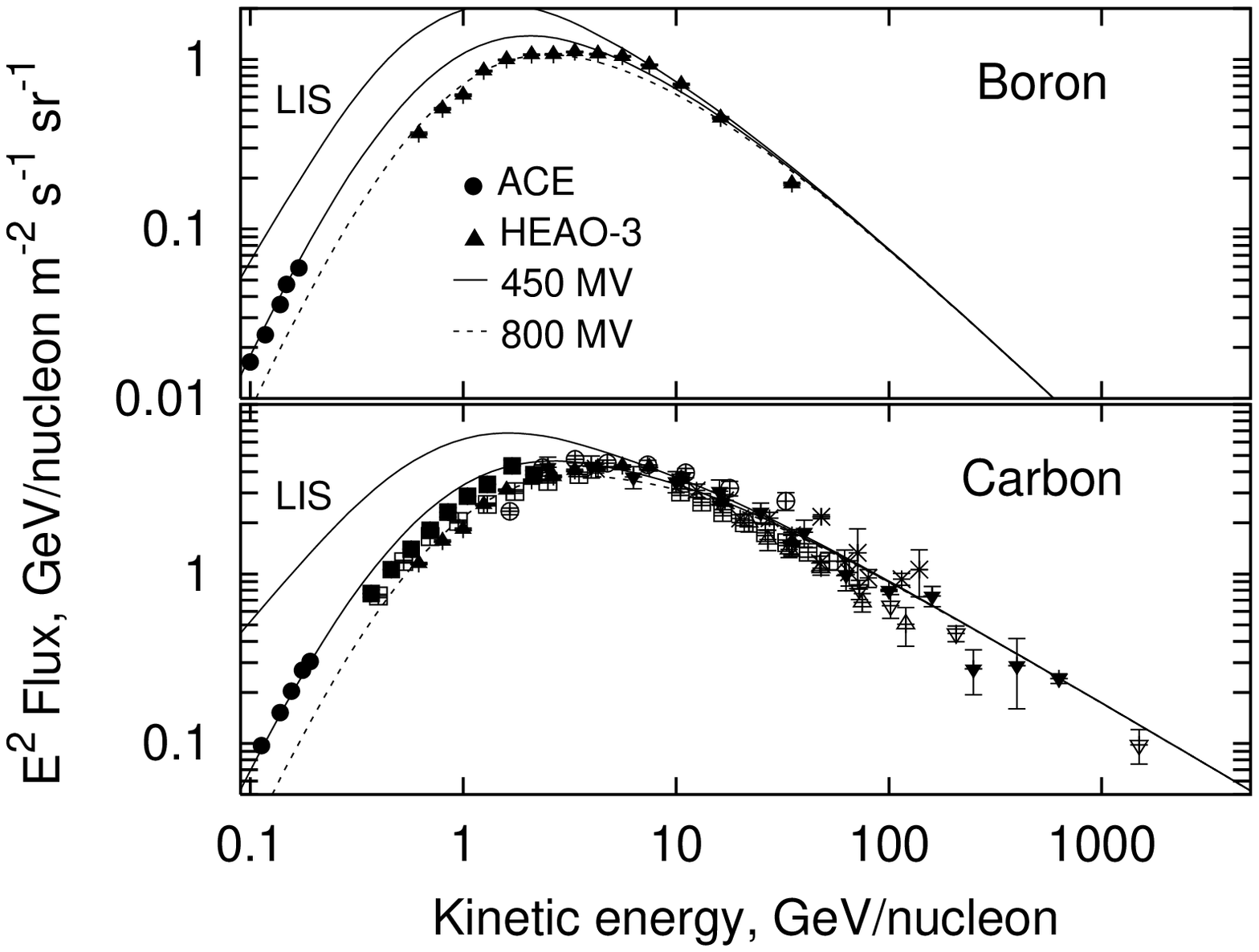}\hfill
\includegraphics[width=\two]{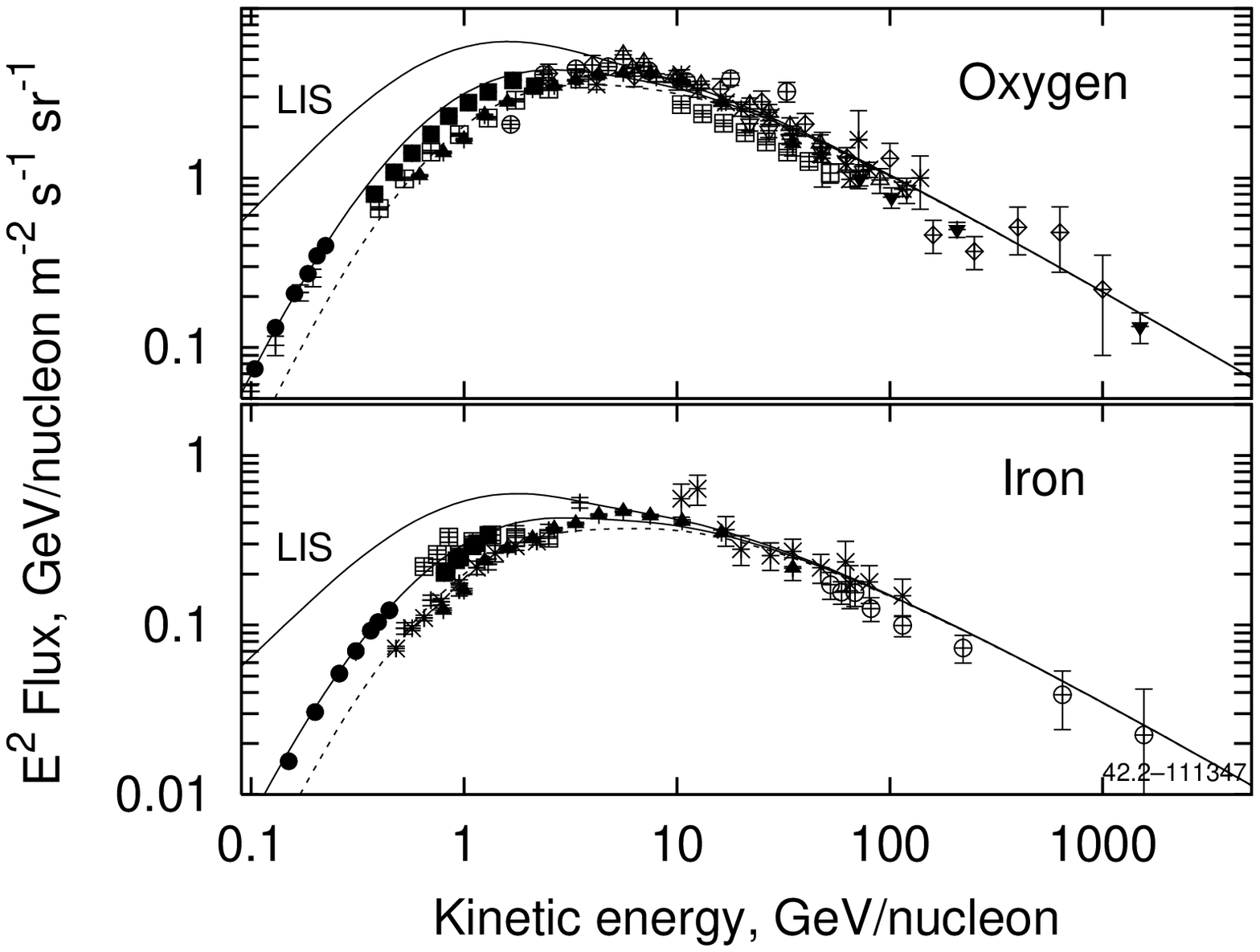}
\end{center}
\vspace{-3\baselineskip}
\caption{\small
Spectra of B, C, O, and Fe 
calculated with LB contribution.
Upper curves - LIS, lower curves -- modulated using force field
approximation ($\Phi=450$ MV -- solid curves, $\Phi=800$ MV -- dashes). 
Data: ACE \citep{davis,davis01},
HEAO-3 \citep{Engelmann90}, for other references see
\citet{StephensStreitmatter98}. \vspace{-0.5\baselineskip}
\label{fig:spectra}}
\end{figure}

Spectra of B, C, O, Fe are shown in Figure~\ref{fig:spectra}
for modulation levels 450 and 800 MV.
In case of C, the normalization coefficient
in the LB component is fixed as $a(6,12)=6.35\times10^{-4}$
cm$^{-2}$ s$^{-1}$ sr$^{-1}$ for $\eta=1$. 
Since the elemental abundances are
tuned at both high and low energies, the calculated sub-Fe/Fe ratio
also agrees well with data.

Secondary nuclei K, Sc, Ti, V 
appear to be overabundant in the LB sources relative to the solar system
(shown as upper limits in Figure~\ref{fig:Ab-ratio}), though the derived
absolute LB abundances are not large. The derived LB abundance  
of Ti does not exceed that of Cr while the derived
abundances of Sc and V are not larger than that of Mn.
One possible reason for this excess is the uncertainty in
the production cross sections, which is especially large for these nuclei. 
Sometimes there is no measurement at all, while phenomenological
systematics are frequently wrong by a factor of two or more.
Often, there is only one measurement at $\sim600$ MeV/n,
which has to be extrapolated in both directions.
This allows only a nearly flat Webber-type or Silberberg-Tsao-type
extrapolations while the real cross sections
usually have large resonances below several hundred MeV, and 
decrease with energy above a few GeV \citep[see, e.g.,][]{MMS}.
We note that \citet{davis} used semiempirical cross sections based on
\citet{W-code} and also predicted fluxes of sub-Fe elements which are too low.  

An estimate of the overall error, which is reflected in the derived LB source
abundances, can be obtained by assuming the absence of 
K, Sc, Ti, V 
in the LB source. Such a discrepancy is below 20\%, and
can be \emph{removed} by allowing the production cross 
sections to increase at low energies by $\sim15-20$\%, which seems plausible.
Another possibility is errors in flux measurements of the rare CR species.
Ulysses and ACE measurements are not always in agreement.
Note that even for such an abundant nucleus as Fe, which is the
main contributor to the sub-Fe group, the discrepancy 
exceeds 10\%, while the disagreement in abundance of Sc is 
$\sim30$\%.

The derived source overabundance of sub-Fe elements in the LB could 
also in principle arise from composition differences between the ISM in the LB and
solar or Galactic average ISM. This is suggested by 
the fact that the relative abundances of secondary elements in the LB
sources are systematically larger than in the Galactic
CR sources (Figure~\ref{fig:Ab-ratio}).

\subsection*{Isotopic distributions in cosmic rays}\label{sec:isotopes}
Be and B isotopes are assumed all secondary, thus there is no
possibility to tune them. Our calculation shows perfect agreement
with the data on relative isotopic abundances of Be and B. 
This is in contrast with a ``standard''
reacceleration model, where we obtained a 15\% discrepancy with relative
abundances of Be and Be isotopes \citep{SM01}.
Abundances of stable isotopes of other elements are not very conclusive
because they are present in the sources, but O and Si isotopic distributions
still agree very well with data by ACE, Ulysses, and Voyager
assuming only \e{16}{}O and \e{28}{}Si isotopes are present in the LB component.
C and N isotopic distributions
do not agree too well, but this may point to a problem with
cross sections.
The calculated ratio \e{13}{}C/\e{12}{}C $\sim0.11$ at 120 MeV/n
($\Phi= 500$ MV) in the model with LB contribution
is still a factor $\sim1.5$ too large compared to the measured value, $0.0629\pm0.0033$
\citep[Voyager 50--130 MeV/n,][]{w96} and 
$0.078\pm0.011$ 
\citep[Ulysses 100--200 MeV/n,][]{d96}, which may 
be connected in part with overproduction of~ \e{13}{}C by \e{15}{}N. 
If we replace the cumulative cross section \e{15}{}N$+p\to$\e{13}{}C
with cross section \e{14}{}N$+p\to$\e{13}{}C, the calculated ratio 
\e{13}{}C/\e{12}{}C will be lowered by 10\% as estimated.
Assuming the absence of the isotope \e{13}{}C
in the Galactic CR sources gives another 10\% reduction. Altogether
these corrections yield \e{13}{}C/\e{12}{}C $\sim0.09$, close to the data.

In case of radioactive isotopes Be, Al, Cl, usually used as ``clocks'' in CR,
the agreement with the most accurate low energy data by ACE
is very good and all the ratios are consistent with each other
indicating that a halo height $z_{\ h} = 4$ kpc is a good estimate. 
Higher energy data by ISOMAX (Be) are also consistent with 
calculations considering the large error bars.
The \e{54}{}Mn/Mn ratio indicates a somewhat smaller halo, but this may be related
to uncertainty in its half-life and/or production cross section. 
The half-life of~ \e{54}{}Mn
against $\beta^-$ decay is the most uncertain among the four
radioactive isotopes -- it is only one which is not measured
directly.
Apart from the half-life, a possible source of errors
can be production cross sections of Mn isotopes.
Only the reaction~ \e{nat}{}Fe$+ p\to$\e{54}{}Mn on a natural sample
of Fe has been measured well enough \citep[see compilation by][]{t16lib}.

\section*{CONCLUSION}\label{sec:conclusion}
In a previous paper we have shown that 
the \antiproton\ flux and B/C ratio appear to
be inconsistent with measurements when computed in standard 
diffusion/reaccelation models.
In this paper we have demonstrated that this discrepancy can be resolved if
some part of the local CR consists of a 
``fresh'' component accelerated in the LB.
The independent evidence for SN activity in the solar vicinity
supports this idea.

Combining the measurements of the \antiproton\ flux \emph{and}
B/C ratio to fix the diffusion coefficient, 
we have been able to construct a model consistent with 
measurements of important nuclei ratios 
in CR and derive elemental abundances in the LB.
Calculated isotopic abundance distributions of Be and B 
are in perfect agreement with CR data.
The abundances of three
``radioactive clock'' isotopes in CR  
\e{10}{}Be, \e{26}{}Al, \e{36}{}Cl
are all consistent and indicate a halo size $z_h\sim4$ kpc based on 
the ACE data.
\e{54}{}Mn indicated a smaller halo, but this may be related
to its half-life uncertainty and/or cross section errors.
The derived fraction of LB component in CR is small compared to
Galactic CR and has a steep spectrum with a cutoff above 
several hundred MeV/n.
Other experimental data (except maybe the overabundance
of Sc, Ti, V) do not contradict this hypothesis. 
The derived source overabundance of sub-Fe elements in the LB 
may be caused by trivial uncertainties in the production cross sections,
or could in principle arise from composition and/or evolution differences between 
the ISM in the LB and solar or Galactic average ISM. 

Recently there has appeared some indication \citep{atmospheric_pbars} that the atmospheric contribution
to the \antiproton\ flux measured in the upper atmosphere is underestimated by $\sim$30\%.
This means that the flux of \antiproton's in CR in reality may be 
\emph{lower} at the top of the atmosphere by at least 25--30\%.
If the latter is true, the reacceleration model (even without LB)
could still be the best one to describe propagation of nucleon species 
in the Galaxy. 

\section*{ACKNOWLEDGEMENTS}

The authors are grateful to M.\ Wiedenbeck for providing the ACE
isotopic abundances. I.\ V.\ M.\ is grateful to the Gamma Ray Group
of the MPI f\"ur extraterrestrische Physik, 
where a part of this work has been done, for hospitality.
I.\ V.\ M.\ and S.\ G.\ M.\ acknowledge partial support from a
NASA ATP grant.

\enlargethispage{3\baselineskip}

\bf
\smallskip\noindent
E-mail address of I.V.~Moskalenko: \hspace{2mm} {imos@milkyway.gsfc.nasa.gov}

\medskip\noindent
Manuscript received \hspace{30mm}; revised \hspace{30mm}; accepted

\end{document}